\documentstyle[12pt]{article}

\font\twlgot =eufm10 scaled \magstep1
\font\egtgot =eufm8
\font\sevgot =eufm7
\font\twlmsb =msbm10 scaled \magstep1
\font\egtmsb =msbm8
\font\sevmsb =msbm7
\newfam\gotfam

\textfont\gotfam\twlgot
\scriptfont\gotfam\egtgot
\scriptscriptfont\gotfam\sevgot

\newfam\msbfam
\textfont\msbfam\twlmsb
\scriptfont\msbfam\egtmsb
\scriptscriptfont\msbfam\sevmsb

\def\pBbb{\relax\ifmmode\expandafter\Bb\else\typeout{You cann't use
Bbb in text mode}\fi}
\def\Bb #1{{\fam\msbfam\relax#1}}
\def\thebibliography#1{\bigskip\section*{\centering
References\\}\bigskip\list
  {\arabic{enumi}.}{\settowidth\labelwidth{#1}\leftmargin\labelwidth
    \advance\leftmargin\labelsep
    \usecounter{enumi}}
    \def\newblock{\hskip .11em plus .33em minus .07em}
    \sloppy\clubpenalty4000\widowpenalty4000
    \sfcode`\.=1000\relax}
\newcommand{\Si}{\Sigma}

\def\op#1{\mathop{\fam0 #1}\limits}

\newcommand{\Id}{{\rm Id\,}}
\def\Ker{{\rm Ker\,}}
\newcommand{\ben}{\begin{eqnarray}}
\newcommand{\een}{\end{eqnarray}}
\newcommand{\be}{\begin{eqnarray*}}
\newcommand{\ee}{\end{eqnarray*}}
\newcommand{\bea}{\begin{eqalph}}
\newcommand{\eea}{\end{eqalph}}
\newcommand{\cL}{{\cal L}}
\newcommand{\cE}{{\cal E}}
\newcommand{\cH}{{\cal H}}
\newcommand{\cF}{{\cal F}}

\newcommand{\al}{\alpha}
\newcommand{\bt}{\beta}
\newcommand{\la}{\lambda}
\newcommand{\La}{\Lambda}

\newcommand{\p}{\pi}
\newcommand{\om}{\omega}
\newcommand{\Om}{\Omega}
\newcommand{\m}{\mu}

\newcommand{\g}{\gamma}
\newcommand{\G}{\Gamma}

\newcommand{\ve}{\varepsilon}
\newcommand{\th}{\theta}

\newcommand{\si}{\sigma}
\newcommand{\w}{\wedge}
\newcommand{\wt}{\widetilde}
\newcommand{\wh}{\widehat}
\newcommand{\ol}{\overline}
\newcommand{\dr}{\partial}

\newcounter{eqalph}
\newcounter{equationa}
\newenvironment{proposition}[1]{{{\it Proposition #1:}}}{}

\newenvironment{definition}[1]{{{\it Definition #1:}}}{}

\newenvironment{eqalph}{\stepcounter{equation}
\setcounter{equationa}{\value{equation}}
\setcounter{equation}{0}

\begin{eqnarray}}{\end{eqnarray}
\setcounter{equation}{\value{equationa}}}

\begin{document}
\hbox{}

\centerline{\bf\large MULTIMOMEMTUM HAMILTONIAN FORMALISM}
\medskip

\centerline{\bf\large IN FIELD THEORY}
\bigskip

\centerline{\bf Gennadi Sardanashvily}
\medskip

\centerline{Department of Theoretical Physics, Moscow State University}

\centerline{117234 Moscow, Russia}

\centerline{E-mail: sard@theor.phys.msu.su}

\begin{abstract}
The standard Hamiltonian machinery, being applied to field theory, leads to
infinite-dimensional phase spaces.
It is not covariant. In this article, we present covariant finite-dimensional
multimomentum Hamiltonian formalism for field theory. This formalism has been
developed from 70th as multisymplectic generalization of the Hamiltonian
formalism in mechanics. In field theory, multimomentum canonical variables
are field functions and momenta corresponding to derivatives of fields with
respect all world coordinates, not only the time. In case of regular
Lagrangian densities, the multimomentum Hamiltonian formalism is equivalent
to the Lagrangian formalism, otherwise for degenerate Lagrangian densities.
In this case, the Euler-Lagrange equations become undetermined and require
additional conditions. In gauge theory, they are gauge conditions. In general
case, these supplementary conditions remain elusive. In the framework of the
multimomentum Hamiltonian machinery, one obtaines them automatically as a
part of Hamilton equations. In case of semiregular and almost regular
Lagrangian densities, we get comprehensive relations between Lagrangian and
multimomentum Hamiltonian formalisms. The key point consists in the fact
that, given a degenerate Lagrangian density, one must consider a family of
associated multimomentum Hamiltonian forms in order to exaust solutions of
the Euler-Lagrange equations. We spell out  degenerate quadratic
and affine Lagrangian densities. The most of  field models are of
these types. As a result, we get the general procedure of describing
constraint field systems.
\end{abstract}

\section{Introduction}

We follow the generally accepted geometric description of classical
fields by sections of fibred manifolds $Y\to X.$
 Lagrangian  and  Hamiltonian formalisms on fibred manifolds
 are phrased in terms of  jet spaces. Given a fibred manifold
$Y\to X$, the $k$-order jet space $J^kY$ of $Y$
comprises the equivalence classes
$j^k_xs$, $x\in X$, of sections $s$ of $Y$ identified by the first $(k+1)$
terms of their Taylor series at a point $x$. One utilizes
 the well-known  facts that  (i) the jet space $J^kY$
 is a finite-dimensional smooth manifold and  (ii)
a $k$-order differential operator on sections of $Y$  can be represented by
a morphism of $J^kY$  to  a vector
bundle over $X$. As a consequence, the dynamics of field systems is
formulated in terms
of finite-dimensional configuration and phase spaces. Moreover,  we
get the  differential geometric description of this dynamics,  for
there is the 1:1 correspondence between sections of the jet bundle $J^1Y\to Y$
and connections on $Y\to X$.

In field theory, we can restrict ourselves
 to the first order  Lagrangian formalism when the configuration space
 is $J^1Y$. Given fibred coordinates $(x^\m, y^i)$
 of $Y$, the jet space $J^1Y$   is endowed with the adapted  coordinates
$ (x^\m, y^i, y^i_\m)$
which bring it into a finite-dimensional smooth manifold.
A first order Lagrangian density is
represented by a horizontal exterior density
 \[
L=\cL(x^\m, y^i, y^i_\m)\om, \qquad \om=dx^1\w ...\w dx^n,
\qquad n=\dim X,
\]
on  $J^1Y\to X$.
The associated Euler-Lagrange
operator $\cE_L$ on the second order jet manifold $J^2Y$ sets up the
  system of second order Euler-Lagrange equations
for sections  of the fibred manifold $Y$. Its canonical extension
$\cE'_L$ to the sesquiholonomic subbundle $\wh J^2Y$ of the repeated jet
manifold $J^1J^1Y$ yields
 the equivalent system of first order Euler-Lagrange equations
for sections $\ol s$ of the fibred jet manifold
$J^1Y\to X$.
 The Poincar\'e-Cartan form
\begin{equation}
\Xi_L=\pi^\la_idy^i\w\om_\la -\pi^\la_iy^i_\la\om +\cL\om, \qquad
\pi^\la_i=\dr^\la_i\cL, \label{303}
\end{equation}
associated with a first order  Lagrangian density $L$ is uniquely defined.
With $\Xi_L$, we have  the  Cartan equations
 \begin{equation}
\ol s^*(u\rfloor d\Xi_L) = 0 \label{316}
\end{equation}
for sections $\ol s$ of  $J^1Y\to X$
where $u$ is an arbitrary  vertical vector field on  $J^1Y\to X$. The equations
(\ref{316}) are
equivalent to the first order Euler-Lagrange equations on sections
$\ol s$ which are jet prolongations  $\ol s=J^1s$ of sections $s$ of $Y\to X$.

At present, we observe the following main  Hamiltonian approaches to field
theory:

(i) the standard Hamiltonian formalism;

 (ii)  the Hamilton-De Donder formalism which  has been developed as
the counterpart of the higher order Lagrangian formalism
in the framework of the calculus of variations [1-6];

 (iii) the multimomentum Hamiltonian formalism which is the multisymplectic
generalization of the conventional Hamiltonian formalism to fibred manifolds
over an $n$-dimensional base $X$, not only ${\bf R}$ [7-12].

 In the straightforward manner when evolution is governed by  the
Poisson bracket, the standart Hamiltonian  formalism
leads to infinite-dimensional
symplectic  spaces \cite{2kos,ber}. Its application to constraint field
systems follows the Dirac's procedure generalized to the
infinite-dimensional case \cite{2got,ber}.
 In the naive Hamiltonian approach to field theory,
 the Hamilton equations are replaced with the
equations of the Hamilton-De Donder type
 with respect to only a time coordinate
$x^0$ by means of substituting solutions $y^i_0$ of the equations
 $p^0_i=\pi^0_i(x^\m,y^i,y^i_\la)$ into the Cartan equations (\ref{316}).

In the calculus of variations,
the phase space is the manifold
\begin{equation}
Z = \op\w^{n-1}T^*X\w T^*Y \label{N41}
\end{equation}
into which the Poincar\'e-Cartan form $\Xi_L$  takes its values
\cite{got,car,2sau}.
  This manifold  is endowed with the coordinates
$(x^\la, y^i, p^\la_i, p)$ such that
\[
(x^\m, y^i, p^\m_i, p)\circ\Xi_L =(x^\m, y^i, \pi^\m_i, \pi^\m_i y^i_\m
-\cL).
\]
It carries  the canonical form
\begin{equation}
\Xi= p\om + p^\la_i dy^i\w\om_\la, \qquad \om_\la = \dr_\la\rfloor\om,
\label{N43}
 \end{equation}
and the corresponding   polisymplectic form  $d\Xi$. In case
of $n=1$, the form $\Xi$  reduces
to the Liouville form $\Xi = Edt + p_i dy^i $
for the homogeneous formalism of mechanics
where $E$ is the energy variable.

In the Hamilton-De Donder approach, a Hamiltonian form fails to be introduced
intrinsically. Given a Lagrangian density $L$, it is defined to be the pullback
of the canonical form (\ref{N43}) by the natural
injection $i_L$  of a submanifold $Z_L = \Xi_L(J^1Y)$ into $Z$. The
corresponding equations are the Hamilton-De Donder equations
\begin{equation}
\ol r^*(u\rfloor dH_L)=0 \label{N46}
\end{equation}
for sections $\ol r$ of the fibred manifold $Z_L\to X$
where $u$ is an arbitrary vertical vector field on  $Z_L\to X$.
When the Poincar\'e-Cartan morphism $\Xi_L: J^1Y\to Z$
is almost regular, the Hamilton-De Donder equations (\ref{N46}) are equivalent
to the Cartan equations (\ref{316}) \cite{got}.
Consequently, in comparison with the Lagrangian machinery, the
Hamilton-De Donder formalism takes no advantageous of describing constraint
systems.

In the multimomentum Hamiltonian formalism,
the phase space is the Legendre manifold
\begin{equation}
\Pi=\op\w^n T^*X\op\otimes_Y TX\op\otimes_Y V^*Y \label{00}
\end{equation}
into which the Legendre morphism $\wh L$ associated with a Lagrangian density
$L$ on $J^1Y$ takes its values.
This manifold is provided with  the fibred coordinates $(x^\la ,y^i,p^\la_i)$
such that
\[
 (x^\m,y^i,p^\m_i)\circ\wh L=(x^\m,y^i,\pi^\m_i).
\]

The Legendre manifold (\ref{00}) carries the generalized Liouville form
\begin{equation}
 \th =-p^\la_idy^i\w\om\otimes\dr_\la	\label{2.4}
\end{equation}
corresponding to the canonical bundle monomorphism
\[
\th :\Pi\op\to_Y\op\w^{n+1} T^*Y\op\otimes_Y TX.
\]
and the associated multisymplectic form
\begin{equation}
\Om =dp^\la_i\w
dy^i\w\om\otimes\dr_\la. \label{406}
\end{equation}
If $X={\bf R}$, these reproduce respectively the
 Liouville form and the symplectic form in mechanics.

The multimomentum Hamiltonian formalism is phrased intrinsically in terms
of Hamiltonian connections which
play the  role similar Hamiltonian vector fields in the symplectic geometry
\cite{gun,6sar,sard}.
We  say that a connection
$\g$ on the fibred Legendre manifold $\Pi\to X
$ is a Hamiltonian connection if the  form  $\g\rfloor\Om$ is closed.
Then, a  Hamiltonian form $H$	 on $\Pi$ is defined to be an
exterior form  such that
 \begin{equation}
dH=\g\rfloor\Om \label{013}
\end{equation}
for some Hamiltonian connection $\g$. Note
that the manifold $Z$ (\ref{N41}) is the 1-dimensional affine bundle over the
Legendre manifold (\ref{00}). There is  the 1:1 correspondence
between the Hamiltonian forms $H$ on  $\Pi$
and the sections $h$ of the bundle $Z\to\Pi$
so that $H$ are the pullbacks of the
canonical form $\Xi$ on $Z$ by the sections $h$ \cite{car}.

The key point consists in the fact that
 every  Hamiltonian form admits splitting
\begin{equation}
H =p^\la_idy^i\w\om_\la
-p^\la_i\G^i_\la\om -\wt{\cH}_\G\om=p^\la_idy^i\w\om_\la-\cH\om \label{017}
\end{equation}
  where $\G$ is a
connection on the fibred manifold $Y$ \cite{car,4sar,6sar}.
In physical applications, one can consider this splitting
as the workable definition of Hamiltonian forms on $\Pi$ \cite{4sar,3sar}.
Given the  Hamiltonian form $H$ (\ref{017}), the equality
(\ref{013}) comes  to the Hamilton equations
 \bea
&&\dr_\la r^i(x) =\dr^i_\la\cH, \label{3.11a}\\
&& \dr_\la r^\la_i(x)
=-\dr_i\cH \label{3.11b}
 \eea
for  sections
$r$ of the Legendre manifold $\Pi\to X$.

If a Lagrangian density $L$ is hyperregular (i.e. $\wh L$ is a
diffeomerphism), there exists the unique Hamiltonian form $H$ such that
the first order Euler-Lagrange equations and the Hamilton equations are
equivalent.

If $X={\bf R},$ the
multimomentum Hamiltonian formalism reproduces the familiar Hamiltonian
formalism of mechanics.
In this case, we have the well-known bijection between the
Hamiltonian forms and the Hamiltonian connections.
If one deals with constraints,
this bijection  makes necessary  the Dirac-Bergman
procedure of constructing the chain of $k$-class
primary constraints spaces \cite{2got}.

In the  multimomentum Hamiltonian formalism when $n>1$, we
have  a set of Hamiltonian connections for the same
 Hamiltonian form. Therefore, one can
choose solutions of the Hamilton equations living on a constraint space.
Moreover, in case of a degenerate  Lagrangian system when the constraint
space  is  $Q=\wh L(J^1Y)$, we observe a
family of different  Hamiltonian forms $H$ associated with the same
Lagrangian density $L$.

If a Lagrangian density is degenerate,
the system of the Euler-Lagrange equations is
underdetermind and require additional conditions. In gauge theory, they
are gauge conditions. In general case, these gauge-type conditions
 remain elusive. In the framework of
the multimomentum Hamiltonian formalism, we get them
as a part of the Hamilton equations (\ref{3.11a}). Different gauge-type
 conditions correspond to different associated multimomentum Hamiltonian
forms.

We shall restrict our consideration to  (i) semiregular Lagrangian densities
$L$ when the preimage $\wh L^{-1}(p)$ of any point $p\in Q$ is a
connected submanifold of  $J^1Y$ and (ii) almost
regular Lagrangian densities when $J^1Y$ is a fibred manifold over $Q$ which
is an imbedded submanifold  of $\Pi$.
These notions of degeneracy seem most appropriate. Lagrangian densities
of fields are almost always both semiregular and almost regular.
In this case, we get the comprehensive relation between
solutions of the Euler-Lagrange equations and the Hamilton equations
\cite{6sar,zak}.

If a Lagrangian density $L$ is semiregular, all associated
Hamiltonian forms $H$ consist with each other on the constraint space $Q$.
For an associated  Hamiltonian form $H$, every solution of the
corresponding Hamilton equations which lives on the constraint space $Q$ yields
a solution of the Euler-Lagrange equations. One
may hope that, conversely, for any solution of the Euler-Lagrange equations,
there exists the corresponding solution of the Hamilton equations for some
associated Hamiltonian form. In case of an almost regular Lagrangian
density which is semiregular, such a complete family of local
Hamiltonian forms always exists.

In Section 6, we spell out models with affine and
almost regular quadratic Lagrangian densities.
 In this case, a complete family of
associated Hamiltonian forms always exists. The
corresponding Hamilton equations are
separated in the dynamic equations and the gauge-type conditions
independent of momenta $p^\la_i$.
As a result, we get the universal procedure of describing constraint
field theories. We apply it to the gauge theory of principal
connections and the gravitation theory in Palatini variables.

\section{Technical preliminary}

 All maps  throughout are  of
class $C^\infty$ and manifolds are real, Hausdorff,
finite-dimensional, second-countable and connected.

A fibred manifold is defined to be the  surjective submersion $\pi:Y\to X$.
A locally trivial fibred manifold is called a bundle.
 By $VY$ and $V^*Y$, we denote  vertical tangent and
vertical cotangent bundles of $Y$ respectively.

Given  fibred manifolds
 $Y\to X$ and  $ Y'\to X'$, let $\Phi : Y\to Y'$
be a fibred manifold morphism over
$ f: X\to X'$. If
$f=\Id_X,$ we  call $\Phi$ the fibred morphism $Y\to_XY'$ over $X$.

Given a fibred manifold $Y\to X$ and a manifold morphism
$f : X' \to X, $ by $f^*Y$ is meant the pullback of $Y$ by $f$  over $X'$.
  For the sake of simplicity,  the pullbacks
$\pi^*(TX)$ and $\pi^*(T^*X)$
are denoted by $TX$ and $T^*X$ respectively.

On fibred manifolds, we
consider  the following special	types of differential forms:

 (i) exterior horizontal forms $ Y\to\op\w^r T^*X;$

(ii) tangent-valued horizontal forms
\be
 && \phi : Y\to\op\w^r T^*X\op\otimes_Y TY,\\
&& \phi =dx^{\la_1}\wedge\dots\wedge dx^{\la_r}\otimes
(\phi_{\la_1\dots\la_r}^\m\dr_\m +
\phi_{\la_1\dots\la_r}^i \dr_i);
\ee
and, in particular, soldering forms
$\si^i_\la dx^\la\otimes\dr_i$;

(iii) pullback-valued forms
 \[
Y\to \op\w^r T^*Y\op\otimes_Y TX, \qquad
Y\to \op\w^r T^*Y\op\otimes_Y V^*Y.
\]

Horizontal $n$-forms are called horizontal densities.

  Given a fibred manifold $Y\to X$, the first order jet manifold $J^1Y$ of
$Y$ is both the fibred manifold $J^1Y\to X$
 and the affine bundle $J^1Y\to Y $  modelled on the vector
bundle  $T^*X\otimes_Y VY. $
 The adapted coordinates $(x^\la, y^i, y^i_\la)$ of $J^1Y$
are compatible with these fibrations:
 \[
 x^\la \to
{x'}^\la(x^\m),\qquad y^i \to {y'}^i(x^\m,y^j), \qquad
  {y'}^i_\la = (\frac{\dr {y'}^i}{\dr y^j}y_\m^j +
\frac{\dr {y'}^i}{\dr x^\m})\frac{\dr x^\m}{\dr {x'}^\la}.
\]
There is the  canonical bundle monomorphism (the contact map)
\[
 \la:J^1Y\op\to_Y
T^*X \op\otimes_Y TY, \qquad \la=dx^\la \otimes
(\dr_\la + y^i_\la \dr_i).
\]

Let  $\Phi$ be a fibred morphism  of $Y\to X$ to
 $Y'\to X$ over a diffeomorphism of $X$.
Its jet prolongation reads
\[
J^1\Phi : J^1Y\to  J^1Y', \qquad {y'}^i_\mu\circ
J^1\Phi=(\dr_\la\Phi^i+\dr_j\Phi^iy^j_\la)\frac{\dr x^\la}{\dr {x'}^\mu}.
\]
The jet prolongation of a section $s$ of $Y\to X$
is the section $y_\la^i\circ J^1s = \dr_\la s^i$
of $J^1Y\to X$.

The repeated jet manifold
$J^1J^1Y$, by definition, is the first order jet manifold of
$J^1Y\to X$. It is provided with the adapted coordinates
$(x^\la ,y^i,y^i_\la ,y_{(\m)}^i,y^i_{\la\m})$.
Its subbundle $ \wh J^2Y$ with $y^i_{(\la)}= y^i_\la$ is called the
sesquiholonomic jet manifold.
The second order jet manifold $J^2Y$ of $Y$ is the subbundle
of $\wh J^2Y$ with $ y^i_{\la\m}= y^i_{\m\la}.$

Given a fibred  manifold  $Y\to X$, a jet field
 $\G$ on $Y$ is defined to be a  section
of the jet bundle $J^1Y\to Y$. A global jet field is a connection on $Y$.
By means of the contact map $\la$, every connection $\G$ on
 $Y$ can be represented by the tangent-valued form
$\la\circ\G$ on $Y$. For the sake of
simplicity, we  denote this
form by the same symbol
\[
\G =dx^\la\otimes(\dr_\la +\G^i_\la\dr_i).
\]

 Let $\G$ be a connection and $\si$ be a
soldering form on a fibred manifold $Y$. Then, their affine sum
$\G +\si$ is a connection on $Y$.
 Let $\G$ and $\G '$ be connections on $Y$.
Then, their affine difference
$\G'-\G$ is a soldering form on  $Y$.

The Legendre manifold  $\Pi$ (\ref{00}) of a fibred manifold $Y$ is the
composite  manifold
\[
\pi_{\Pi X}=\pi\circ\pi_{\Pi Y}:\Pi\to Y\to X
\]
 endowed with the fibred coordinates  $( x^\la ,y^i,p^\la_i)$:
\[
 {p'}^\la_i = J \frac{\dr y^j}{\dr{y'}^i} \frac{\dr
{x'}^\la}{\dr x^\m}p^\m_j, \qquad J^{-1}=\det (\frac{\dr {x'}^\la}{\dr
x^\m}) .
\]
By $J^1\Pi$ is meant the first order jet manifold of  $\Pi\to X$.
It is provided with the adapted fibred coordinates
$
( x^\la ,y^i,p^\la_i,y^i_{(\m)},p^\la_{i\m}) .
$

By a momentum morphism, we call a fibred morphism
\[
\Phi: \Pi\op\to_Y J^1Y, \qquad ( x^\la ,y^i,y^i_\la)\circ\Phi= ( x^\la
,y^i,\Phi^i_\la(p)).
 \]
 Given a momentum morphism $\Phi$, its composition with the
contact map $\la$
is represented by the  horizontal pullback-valued 1-form
\begin{equation}
\Phi =dx^\la\otimes(\dr_\la +\Phi^i_\la (p)\dr_i)\label{2.7}
 \end{equation}
on $\Pi\to X$.
For instance, let $\G$ be a connection on $Y\to X$. Then,
$ \wh\G=\G\circ\pi_{\Pi Y}$
 is a momentum morphism.  Conversely,
every momentum morphism $\Phi$ of the Legendre manifold $\Pi$ of $Y$ defines
the associated connection $ \G_\Phi =\Phi\circ\wh 0$
on $Y\to X$ where $\wh 0$ is the global zero section of the
Legendre bundle $\Pi\to Y$.

\section{Lagrangian formalism}

Given  a  Lagrangian density  $L$,  the jet manifold
$J^1Y$  is provided with the Lagrangian  multisymplectic form
\[
\Om_L=(\dr_j\pi^\la_idy^j+\dr^\m_j\pi^\la_idy^j_\m)\w
dy^i\w\om\otimes\dr_\la.
\]
Using the pullback of this form and the Poincar\'e-Cartan form $\Xi_L$
(\ref{303}) to the
repeated jet manifold $J^1J^1Y$,
	 one can construct the exterior form
\begin{equation}
\La_L=d\Xi_L-\la\rfloor\Om_L
 =(y^i_{(\la)}-y^i_\la)d\pi^\la_i\w\om +
 (\dr_i-\wh\dr_\la\dr^\la_i)\cL dy^i\w\om,\label{304}
\end{equation}
\[
\la=dx^\la\otimes\wh\dr_\la,
\qquad
\wh\dr_\la =\dr_\la +y^i_{(\la)}\dr_i+y^i_{\m\la}\dr^\m_i,
\]
on $J^1J^1Y$.
Its restriction to the second order jet manifold $J^2Y$  reproduces
the familiar variational Euler-Lagrange operator
\ben
&&\cE_L: J^2Y\to\op\w^{n+1}T^*Y,\nonumber \\
&&\cE_L= (\dr_i-\wh\dr_\la\dr^\la_i)\cL dy^i\w\om ,
\qquad \wh\dr_\la =\dr_\la +y^i_\la\dr_i+y^i_{\m\la}\dr^\m_i.
\label{305}
\een
The restriction of the form (\ref{304}) to the sesquiholonomic jet manifold
$\wh J^2Y$ of $Y$ defines the sesquiholonomic extension
\begin{equation}
\cE'_L:\wh J^2Y\to\op\w^{n+1}T^*Y \label{2.26}
\end{equation}
 of the Euler-Lagrange operator (\ref{305}).
 It has the form
(\ref{305}) with nonsymmetric coordinates $y^i_{\m\la}$.

 Let $\ol s$ be a section of the fibred jet manifold $J^1Y\to X$
such that its first order jet prolongation $J^1\ol s$ takes its values into
$\Ker\cE'_L$. Then, it satisfies the system of first order
Euler-Lagrange equations
 \begin{equation}
\dr_\la\ol s^i=\ol s^i_\la, \qquad
 \dr_i\cL-(\dr_\la+\ol s^j_\la\dr_j
+\dr_\la\overline s^j_\m\dr^\m_j)\dr^\la_i\cL=0. \label{306}
\end{equation}
They are equivalent to the  familiar second order
Euler-Lagrange equations
\begin{equation}
\cE_L\circ J^2s =0 \label{2.29}
 \end{equation}
for sections $s$ of $Y\to X$. We have $\ol s=J^1s$.

\section{Multimomentum  Hamiltonian formalism}

Let $\Pi$ be the Legendre manifold (\ref{00}) provided with the generalized
Liouville form $\theta$ (\ref{2.4}) and the multisymplectic form $\Om$
(\ref{406}).

 \begin{definition}{1} We say that a  jet field  (resp. a connection)
\[
\g =dx^\la\otimes(\dr_\la +\g^i_{(\la)}\dr_i
+\g^\m_{i\la}\dr^i_\m)
\]
on the Legendre manifold $\Pi\to X$ is a Hamiltonian jet field
(resp. a Hamiltonian connection) if the exterior form
\[
\g\rfloor\Om =dp^\la_i\w dy^i\w\om_\la
+\g^\la_{i\la}dy^i\w\om -\g^i_{(\la)} dp^\la_i\w\om
\]
 is closed.
\end{definition}

 Hamiltonian connections
constitute an affine subspace of connections on
$\Pi\to X$. The following construction  shows that this subspace is
not empty.

 Every connection $\G$ on  $Y\to X$
is lifted to the connection
\[
 \g=\wt\G =dx^\la\otimes[\dr_\la +\G^i_\la (y)\dr_i	 +
(-\dr_j\G^i_\la (y)  p^\m_i-K^\m{}_{\nu\la}(x) p^\nu_j+K^\al{}_{\al\la}(x)
p^\m_j)\dr^j_\m]
 \]
  on  $\Pi\to X$ where $K$  is a linear
symmetric connection  on the bundles $TX$ and $T^*X$.
We have the equality
\begin{equation}
\wt\G\rfloor\Om =d(\wh\G\rfloor\th) \label{412}
\end{equation}
which shows that  $\wt\G$ is a Hamiltonian connection.

\begin{definition}{2} An exterior $n$-form $H$ on the
Legendre manifold $\Pi$ is called a  Hamiltonian form if, on an
open neighborhood of each point of $\Pi$,
 there exists a Hamiltonian jet field satisfying the equation
$\g\rfloor\Om  =dH.$
 This jet field $\g$ is termed the Hamiltonian jet field for  $H$.
 \end{definition}

Note that   Hamiltonian forms throughout   are considered modulo
closed forms since closed forms do not make any contribution in the Hamilton
equations.

\begin{proposition}{3} Let $H$ be a Hamiltonian form. For
any exterior horizontal density
$\wt H=\wt{\cH}\om$
 on the fibred Legendre manifold $\Pi\to X$, the form $H-\wt H$
is a Hamiltonian form.
Conversely, if $H$ and $H'$ are  Hamiltonian forms,
their difference $H-H'$ is an exterior horizontal density
 on $\Pi\to X$.
\end{proposition}

In virtue of  Proposition 3,  Hamiltonian  forms constitute an affine space
modelled on a linear space of the exterior horizontal densities on
$\Pi\to X$.
A glance
at the equality (\ref{412}) shows that this affine space is not empty.
Given a connection $\G$ on a  $Y\to X$, its lift $\wt\G$ on
 $\Pi\to X$  is a Hamiltonian connection for the Hamiltonian form
\begin{equation}
 H_\G=\wh\G\rfloor\th =p^\la_i dy^i\w\om_\la -p^\la_i\G^i_\la (y)\om.
\label{3.6}
\end{equation}
It follows that every
Hamiltonian form on the Legendre manifold $\Pi$ can be given by the expression
(\ref{017}).

Moreover, a Hamiltonian form has the canonical splitting (\ref{017})
as follows.  Every momentum morphism $\Phi$ represented by
the pullback-valued form (\ref{2.7}) on $\Pi$ yields the associated
 Hamiltonian form
\begin{equation}
H_\Phi=\Phi\rfloor\th =p^\la_idy^i\w\om_\la -p^\la_i\Phi^i_\la\om.
\label{414}
\end{equation}
Conversely, every  Hamiltonian form $H$
 defines the associated momentum morphism
\[
 \wh H:\Pi\to J^1Y, \qquad y_\la^i\circ\wh H=\dr^i_\la\cH,
\]
and the associated connection
$\G_H =\wh H\circ\wh 0$
on $Y\to X$. As a consequence, we have the canonical splitting
\begin{equation}
H=H_{\G_H}-\wt H.\label{3.8}
\end{equation}

\begin{definition}{4} The
Hamilton operator $\cE_H$ for  a  Hamiltonian form $H$
 is defined to be the first order differential operator
\ben
&& \cE_H :J^1\Pi\to\op\w^{n+1} T^*\Pi,\nonumber \\
&& \cE_H=dH-\wh\Om=[(y^i_{(\la)}-\dr^i_\la\cH) dp^\la_i
-(p^\la_{i\la}+\dr_i\cH) dy^i]\w\om \label{3.9}
\een
where
\[
\wh\Om=dp^\la_i\w
 dy^i\w\om_\la +p^\la_{i\la}dy^i\w\om -y^i_{(\la)} dp^\la_i\w\om
\]
 is the pullback of the multisymplectic form $\Om$ onto $J^1\Pi$.
\end{definition}

 For any jet field $\g$ on the Legendre manifold $\Pi$, we have
\[
\cE_H\circ\g =dH-\g\rfloor\Om.
\]
 It follows that  $\g$  is a Hamiltonian jet field for a
 Hamiltonian form $H$ if and only if it takes its values into
$\Ker\cE_H$, that is, satisfies  the algebraic Hamilton equations
\begin{equation}
 \g^i_\la =\dr^i_\la\cH, \qquad
 \g^\la_{i\la}=-\dr_i\cH. \label{3.10}
\end{equation}
In particular,  Hamiltonian jet fields associated with the same
 Hamiltonian form differ from each other in soldering forms
$\wt\si$ on $\Pi\to X$ which obey the relation
\begin{equation}
\wt\si\rfloor\Om =0, \qquad
\wt\si^i_\la =0,\qquad \wt\si^\la_{i\la} =0. \label{2.9}
\end{equation}

 The Hamilton operator (\ref{3.9}) is an affine	morphism and
$ \Ker\cE_H\to\Pi$  is an affine subbundle of
the jet bundle $J^1\Pi\to\Pi$.
Moreover, a glance at the condition (\ref{2.9}) shows that,
for  all Hamilton operators $\cE_H$, this affine subbundle  is
modelled on the same vector subbundle of the bundle $
T^*X\otimes_\Pi V\Pi.$
As a consequence, we have the following assertion.

\begin{proposition}{5} A Hamiltonian connection for every
 Hamiltonian form   always exists.
\end{proposition}

Let  $r$ be a section  of the fibred Legendre manifold $\Pi\to X$ such that
its  jet prolongation $J^1r$ takes its values into $\Ker\cE_H$.
Then, the Hamilton equations (\ref{3.10}) are brought to the first
order differential Hamilton equations  (\ref{3.11a})  and (\ref{3.11b})
Conversely, if  $r$ is a
solution (resp. a global solution) of the Hamilton equations
(\ref{3.11a}) and (\ref{3.11b}) for a
 Hamiltonian form $H$, there exists an extension of this
solution to a Hamiltonian jet field (resp. a Hamiltonian connection) $\g$
which  has an integral section $r$, that is,  $\g\circ r=J^1r$.

\section{Constraint systems}

This Section is devoted to relations between the Lagrangian formalism
on fibred manifolds and
the multimomentum Hamiltonian formalism in case of degenerate Lagrangian
densities.

  Given a fibred manifold $Y\to X$, let
$L$ be a first order Lagrangian density.
One observes that, when the Legendre morphism
$\wh L$ is a diffeomorphism, the corresponding Lagrangian
system 	 meets the unique equivalent Hamiltonian
system. It follows that, if a Legendre morphism
 is regular at a point, the corresponding Lagrangian
system reduced to an open neighborhood of this point has the
equivalent local Hamiltonian system. In order to keep this
equivalence in case of degenerate systems, we require that the image of a
configuration space by the Legendre morphism
contains  all points where the momentum morphism is regular.

\begin{definition}{6} We shall say that a  Hamiltonian form
$H$ is associated with a Lagrangian density $L$ if $H$ satisfies the relations
\bea
&&\wh L\circ\wh H\mid_Q=\Id_Q, \qquad   Q=\wh L( J^1Y) \label{2.30a}, \\
&& H=H_{\wh H}+L\circ\wh H. \label{2.30b}
\eea
\end{definition}

The relation (\ref{2.30b}) results in the condition
\[
(p^\m_i-\dr^\m_i\cL\circ\wh H)\dr^i_\m\dr^a_\al\cH=0.
\]
A glance at this condition  shows that
 (i) the condition (\ref{2.30a}) is the corollary of the
condition (\ref{2.30b}) if the momentum morphism $\wh H$
is regular at all points of $Q$ and
(ii)  $\wh H$ is not regular outside the constraint space $Q$.

Let us emphasize that
different  Hamiltonian forms can be associated with the same
Lagrangian density.

\begin{proposition}{7}
Given a Lagrangian density $L$, let $\Phi$ be a momentum morphism associated
with the Legendre morphism $\wh L$, that is,
 \begin{equation}
\wh L\circ\Phi\mid_Q=\Id_Q. \label{4.26}
\end{equation}
Let us consider the  Hamiltonian form
\begin{equation}
 H_{L\Phi}=H_\Phi+L\circ\Phi\label{4.24}
\end{equation}
where $H_\Phi$ is the Hamiltonian form (\ref{414}).  If
  \begin{equation}
\wh H_{L\Phi} = \Phi, \label{4.25}
\end{equation}
then the  Hamiltonian form (\ref{4.24}) is associated with $L$.
 \end{proposition}

If a Lagrangian density $L$ is hyperregular,
 there always exists the unique  Hamiltonian form
\[
H=H_{\wh L^{-1}}+L\circ\wh L^{-1}
\]
 associated  with $L$.

Contemporary field theories are almost never regular, but semiregular and
almost regular.
Also in this case, one can get the workable relations between
 Lagrangian and multimomentum Hamiltonian formalisms \cite{6sar,sard,zak}.

\begin{proposition}{8} All  Hamiltonian forms associated
with a semiregular Lagrangian density $L$ consists with each other on the
constraint space $Q$,
and the Poincar\'e-Cartan form $\Xi_L$ for $L$ is the pullback
\begin{equation}
 \Xi_L=H\circ\wh L,\qquad
\pi^\la_iy^i_\la-\cL=\cH(x^\m,y^i,\pi^\la_i), \label{2.32}
\end{equation}
 of any associated  Hamiltonian form.
\end{proposition}

\begin{proposition}{9} Let $H$ be a  Hamiltonian form
associated with a semiregular Lagrangian density $L$. The Hamilton
operator $\cE_H$ for $H$ satisfies the relation
\[
\La_L=\cE_H\circ J^1\wh L.
\]
\end{proposition}

 \begin{proposition}{10} (i) Let a  section $r$ of $\Pi\to X$
be a solution of the Hamilton equations (\ref{3.11a}) and (\ref{3.11b})
 for a Hamiltonian form $H$ associated with a semiregular Lagrangian
density $L$. If $r$ lives on the constraint space $Q$, the section
$\ol s=\wh H\circ r$ of $J^1Y\to X$ satisfies the first
order Euler-Lagrange equations (\ref{306}).
(ii) Given a semiregular Lagrangian density $L$, let
 $\ol s$ be a solution of the
first order Euler-Lagrange equations (\ref{306}).
Let $H$ be a Hamiltonian form associated with $L$ so that
\begin{equation}
\wh H\circ \wh L\circ \ol s=\ol s.\label{2.36}
\end{equation}
 Then, the section $r=\wh L\circ \ol s$
 of $\Pi\to X$ is a solution of the
Hamilton equations (\ref{3.11a}) and (\ref{3.11b}) for $H$. It lives on the
constraint space $Q$.
(iii) For every sections $\ol s$ and $r$ satisfying (i) or (ii),
we have the relations
 \[
\ol s=J^1s, \qquad  s= \pi_{\Pi Y}\circ r
\]
where $s$ is a solution of the second order Euler-Lagrange equations
(\ref{2.29}).
\end{proposition}

 We shall say that a family of Hamiltonian forms $H$
associated with a semiregular Lagrangian density $L$ is
complete if, for each solution $\ol s$ of the first order Euler-Lagrange
equations  (\ref{306}), there exists
a solution $r$ of the Hamilton equations (\ref{3.11a}) and (\ref{3.11b}) for
some  Hamiltonian form $H$ from this family so that
\[
 r=\wh L\circ\ol s,\qquad  \ol s =\wh H\circ r, \qquad
 \ol s= J^1(\pi_{\Pi Y}\circ r).
\]
In virtue of Proposition 10, such a complete family
exists if and only if, for each solution $\ol s$ of the Euler-Lagrange
equations for $L$, there exists a  Hamiltonian form $H$ from this
family so that the condition (\ref{2.36}) holds.

We establish existence of a complete family of
Hamiltonian forms associated with a Lagrangian density which is
both semiregular and almost regular \cite{sard,zak}.

\begin{proposition}{11} Let $L$ be an almost regular Lagrangian density.
(i) On an open neighborhood of each point $q\in Q$, there exists a momentum
morphism $\Phi$ which obeys the conditions (\ref{4.26}) and
 (\ref{4.25}) and so, the local  Hamiltonian form
(\ref{4.24}) is associated with $L$.
(ii) If $L$ is a semiregular Lagrangian density,
 there exists a complete family of local associated
Hamiltonian forms
on an open neighborhood of each point $p\in Q$.
\end{proposition}

 Note that, if the imbedded constraint space $Q$ is  not an open
subbundle of $\Pi$, it can not be provided with the
multisymplectic structure.
  Therefore, to consider solutions of the Hamilton
equations   on $Q$, one must introduce a Hamiltonian
form and  the corresponding  Hamilton equations at least on some open
neighborhood of  $Q$. Another way consists in
constructing the Hamilton-De Donder type equations on the imbedded constraint
space $Q$. Let an almost regular Lagrangian density $L$ be semiregular and
$H_Q$ the restriction of the associated  Hamiltonian forms
to $Q$. For sections $r$ of the fibred manifold $Q\to X$,
we can write the equations
\begin{equation}
r^*(u\rfloor dH_Q) =0 \label{N44}
\end{equation}
where $u$ is an arbitrary vertical vector field on $Q\to X$. Since
$dH_Q\neq dH\mid_Q$, these equations fail to be equivalent to the Hamilton
equations restricted to  $Q$. At the same time, we have
$Z_L = H_Q(Q)$ and the equations (\ref{N44}) are equivalent to the
Hamilton-De Donder equations (\ref{N46}).

\section{Quadratic and affine Lagrangian densities}

In this Section,
we  construct complete families of  Hamiltonian forms
associated with  almost regular quadratic and affine
Lagrangian  densities. They are semiregular.

 Let us consider a quadratic Lagrangian density given by the
coordinate expression
 \begin{equation}
 L=\cL\om, \qquad
 \cL=\frac12 a^{\la\m}_{ij}(y)y^i_\la y^j_\m +
b^\la_i(y)y^i_\la + c(y), \label{N12}
\end{equation}
where $a$, $b$ and $c$ are local functions on $Y$ with the corresponding
transformation laws. It is semiregular. The associated Legendre morphism reads
\begin{equation}
p^\la_i\circ\wh L= a^{\la\m}_{ij} y^j_\m +b^\la_i. \label{N13}
\end{equation}
	It is an affine morphism over $Y$. We have the corresponding linear
morphism
\[
\ol L: T^*X\op\otimes_YVY\to\Pi, \qquad
 p^\la_i\circ\ol L
=a^{\la\m}_{ij} y^j_\m.
\]

Note that almost all quadratic Lagrangian densities of field models take the
form
\begin{equation}
 \cL=\frac12 a^{\la\m}_{ij}\ol y^i_\la\ol y^j_\m
 + c(y), \qquad \ol y^i_\m = y^i_\m -\G^i_\m, \label{N15}
\end{equation}
where $\G$ is a connection on $Y$. 	 It is equivalent to the
fact that the constraint space $Q$ given by the Legendre morphism (\ref{N13})
contains the zero section $\wh 0(Y)$ of $\Pi\to Y$.
Then, $\Ker\wh L$ is an affine subbundle of  $J^1Y\to Y$ and
there exists a connection $\G$ on $Y$ 	such that
\begin{equation}
\G: Y\to \Ker\wh L, \qquad
 a^{\la\m}_{ij}\G^j_\m + b^\la_i =0.
\label{N16}
\end{equation}
With this connection, the Lagrangian density (\ref{N12}) can be brought into
the form (\ref{N15}). If the Lagrangian density (\ref{N12}) is regular, the
connection (\ref{N16}) is unique.

 Let $L$ be an almost regular quadratic Lagrangian
density such that $\wh 0(Y)\subset Q$. Then, there exists a linear
pullback-valued horizontal 1-form
\begin{equation}
\si: \Pi\to T^*X\op\otimes_YVY, \qquad
 \ol y^i_\la\circ\si =
\si^{ij}_{\la\m}p^\m_j, \label{N17}
\end{equation}
on $\Pi\to X$ which satisfies the condition
\begin{equation}
\ol L\circ\si\circ i_Q= i_Q \label{N45}
\end{equation}
where $i_Q$ denotes the imbedding of $Q$ into $\Pi$.

The connection (\ref{N16}) and the form (\ref{N17}) are the key objects in
our construction.

Since $\ol L$ and $\si$ are linear morphisms, $\ol L\circ\si$ is a surjective
submersion of $\Pi$ onto $Q$.
It follows that
\begin{equation}
\si=\si\circ\ol L\circ\si, \label{N21}
\end{equation}
and we have the splitting
\begin{equation}
J^1Y=\Ker\wh L\op\oplus_Y{\rm Im}\si. \label{N18}
\end{equation}
 Moreover, there exists $\si$ such that
\begin{equation}
\Pi=\Ker\si \op\oplus_Y Q. \label{N20}
\end{equation}

Given a form $\si$ (\ref{N17}) and a connection (\ref{N16}), let us consider
the affine momentum morphism
\begin{equation}
\Phi=\G+\si,\qquad
 \Phi^i_\la = \G^i_\la (y) + \si^{ij}_{\la\m}p^\m_j.
\label{N19}
\end{equation}
It is associated with the Legendre morphism (\ref{N13}). Conversely, every
affine momentum morphism associated with (\ref{N13}) is of the type
(\ref{N19}). The  Hamiltonian form $H_{L\Phi}$ (\ref{4.24})
corresponding to $\Phi$ is associated
 with the Lagrangian density (\ref{N12})
 due to the condition (\ref{N21}). It reads
\begin{equation}
H= p^\la_idy^i\w\om_\la - (  \G^i_\la
(p^\la_i-\frac12 b^\la_i) + \frac12 \si^{ij}_{\la\m}p^\la_ip^\m_j  -c)\om.
\label{N22}
\end{equation}
The canonical splitting (\ref{3.8}) of this  Hamiltonian form
shows that $\G_H=\G$ and the Hamiltonian density $\wt H$ is quadratic, but it
becomes affine on $\Ker\si$.

We aim to show that the  Hamiltonian forms (\ref{N22}) for
different connections (\ref{N16}) constitute the complete family.

Let us consider
the Hamilton equations (\ref{3.11a}) for the  Hamiltonian
form (\ref{N22}). For sections $r$ of $\Pi\to X$, they read
\begin{equation}
J^1s= (\G+\si)\circ r, \qquad s=\pi_{\Pi Y}. \label{N29}
\end{equation}
With splitting (\ref{N18}), we have the following two surjections
\be
&&{\cal S}: J^1Y\to\Ker\wh L, \qquad
 {\cal S}: y^i_\la\to y^i_\la
-\si^{ik}_{\la\al} (a^{\al\m}_{kj}y^j_\m + b^\al_k), \\
&&\cF: J^1Y\to{\rm Im}\si, \qquad
 \cF=\si\circ\wh L:
y^i_\la\to \si^{ik}_{\la\al} (a^{\al\m}_{kj}y^j_\m + b^\al_k).
\ee
With respect to these projections, the Hamilton equations (\ref{N29}) are
brought into  the pair of equations
\ben
&&{\cal S}\circ J^1s=\G\circ s, \label{N23}\\
&&\cF \circ J^1s=\si\circ r. \label{N28}
\een

On an analogy with gauge theory, by a gauge-type class is meant the preimage
${\cF}^{-1}(\ol y)$ of every point $\ol y\in{\rm Im}\si$.
 Then, the equations (\ref{N23})
make the sense of  gauge-type conditions. We say that these conditions are
universal if they single out one representative of every gauge-type class.
It is
readily observed that the conditions (\ref{N23}) are universal gauge-type
conditions on sections of the jet bundle $J^1\Pi\to\Pi$ when the algebraic
Hamilton equations (\ref{3.10}) are considered, otherwise on sections $r$ of
$\Pi\to X$.
At the same time, they show that  it is condition on ${\cal S}\circ J^1s$
 which can supplement the underdetermined Euler-Lagrange
equations for the degenerate Lagrangian density (\ref{N12}).
Moreover, for every section $s$ of $Y$, there exists a connection $\G$
(\ref{N16}) such that the gauge-type conditions (\ref{N23}) hold.
In this case, the momentum morphism (\ref{N19}) satisfies the condition
\[
\Phi\circ\wh L\circ J^1s= J^1s.
\]
Hence, the Hamiltonian forms (\ref{N22}) constitute a complete family.
Note that  Hamiltonian forms from this family differ
only in connections $\G$ which are responsible for the gauge-type condition
(\ref{N23}).

Let us consider now an affine Lagrangian density
\begin{equation}
L=\cL\om, \qquad \cL=b^\la_i(y)y^i_\la + c(y). \label{N24}
\end{equation}
It is almost regular and semiregular. The corresponding Legendre morphism
reads
\begin{equation}
p^\la_i\circ\wh L = b^\la_i(y). \label{N25}
\end{equation}
It follows that, in case of an affine Lagrangian density, the corresponding
constraint space is given by the section $b$ of  $\Pi\to Y$.

It is easy to see that the  Hamiltonian form
\begin{equation}
H=H_\G+L\circ \G=p^\la_idy^i\w\om_\la - (p^\la_i -b^\la_i)\G_\la^i\om +
c\om,  \label{N26}
\end{equation}
for any connections $\G$ on  $Y$ is associated with $L$
(\ref{N24}). The corresponding momentum morphism reads
\begin{equation}
y^i_\la\circ\wh H = \G^i_\la. \label{N27}
\end{equation}
We thus observe that, in case of the affine Hamiltonian density, the Hamilton
equations (\ref{3.11a}) reduce to the gauge-type conditions (\ref{N27}).

 For any section $s$ of $Y$, we can choose the connection $\G$
which has the integral section $s$. Then, the corresponding momentum morphism
(\ref{N27}) obeys the condition \[
\wh H\circ\wh L\circ J^1s = J^1s.
\]
It follows that the  Hamiltonian forms (\ref{N26}) constitute
the complete family.

\section{Gauge theory of principal connections}

In this Section, the manifold $X$ is assumed to be oriented. It is
provided with a nondegenerate fibre metric $g_{\m\nu}$ and
 $g^{\m\nu}$ in the tangent and cotangent bundles of $X$. We
denote $g=\det(g_{\m\nu}).$

 Gauge theory of
principal connections is described by the degenerate quadratic Lagrangian
density, and its multimomentum Hamiltonian formulation exemplifies
 the common attributes of the	 degenerate quadratic models. The feature of
gauge theory consists  in the fact that  splittings (\ref{N18}) and
(\ref{N20}) are canonical.

Let $P\to X$ be a principal bundle with a structure Lie group $G$
wich acts on $P$ on the right by the law
\[
r_g: P\to Pg, \qquad g\in G.
\]
A principal connection is defined to be a
 $G$-equivariant global jet field $A$ on $P$:
 \[
A\circ r_g=J^1r_g\circ A, \qquad g\in G.
\]

There is the 1:1 correspondence between the principal connections $A$ on
 $P$  and the global sections of the bundle $C=J^1P/G$ called
 the bundle of principal connections. It is the affine bundle modelled on
the vector bundle
 \[
\ol C =T^*X \otimes V^GP, \qquad  V^GP=VP/G.
\]

Given a bundle atlas $\Psi^P$ of $P$, the bundle $C$
is provided with  the fibred coordinates $(x^\mu,k^m_\mu)$ so that
\[
(k^m_\mu\circ A)(x)=A^m_\mu(x)
\]
 are coefficients of the local connection 1-form
of a principal connection
$A$ with respect to the atlas $\Psi^P$.

The first order jet manifold $J^1C$ of the bundle $C$
 provided with the adapted coordinates $
(x^\mu, k^m_\mu, k^m_{\mu\lambda}).$
There exists the canonical splitting \cite{2man}
\begin{equation}
J^1C=C_+\op\oplus_C C_-=(J^2P/G)\op\oplus_C
(\op\wedge^2 T^*X\op\otimes_C V^GP), \label{N31}
\end{equation}
\[
 k^m_{\mu\lambda}=\frac12(
k^m_{\mu\lambda}+k^m_{\lambda\mu} +c^m_{nl}k^n_\lambda k^l_\mu)
 +\frac12( k^m_{\mu\lambda}-k^m_{\lambda\mu}
 -c^m_{nl}k^n_\lambda k^l_\mu),
\]
over $C$ where $C_+\to C$ is the affine bundle modelled on the vector bundle
\[
\ol C_+=\op\vee^2 T^*X\op\otimes_C V^GP.
\]

 There are  the corresponding canonical
surjections:

(i) ${\cal S}: J^1 C\to C_+.$

(ii) $ \cF: J^1 C\to C_-$ where
\[
\cF=\frac{1}{2}\cF^m_{\la\m}dx^\la\wedge dx^\m\otimes e_m,
\qquad \cF^m_{\lambda\mu}=
k^m_{\mu\lambda}-k^m_{\lambda\mu} -c^m_{nl}k^n_\lambda k^l_\mu,
\]

 The Legendre manifold of the bundle $C$ of principal connections reads
\[
\Pi=\op\wedge^n T^*X\otimes TX\op\otimes_C [C\times\ol C]^*.
\]
It is provided with the fibred coordinates
$(x^\mu,k^m_\mu,p^{\mu\lambda}_m) $
and has the canonical splitting
\begin{equation}
\Pi=\op\wedge^n T^*X\op\otimes_C [\ol C^*_+
\op\oplus_C C^*_-]=\Pi_+\op\oplus_C\Pi_-,\label{N32}
\end{equation}
\[
 (k^m_\mu,p^{\mu\lambda}_m)=
(k^m_\mu,p^{(\mu\lambda)}_m=\frac{1}{2}[p^{\mu\lambda}_m+
p^{\lambda\mu}_m]) +(k^m_\mu,
 p^{[\mu\lambda]}_m=\frac{1}{2}[p^{\mu\lambda}_m-
p^{\lambda\mu}_m]).
\]

 On the configuration space (\ref{N31}),
the conventional Yang-Mills Lagrangian density $L_{YM}$
is given by the expression
\begin{equation}
L_{YM}=\frac{1}{4\ve^2}a^G_{mn}g^{\lambda\mu}g^{\beta\nu}\cF^m_{\lambda
\beta}\cF^n_{\mu\nu}\sqrt{\mid g\mid}\,\omega \label{5.1}
\end{equation}
where  $a^G$ is a nondegenerate $G$-invariant metric
  in the Lie algebra of $G$. It is almost regular and semiregular.
The Legendre morphism associated with the Lagrangian density (\ref{5.1}) takes
the form
 \bea
&& \wh L_{YM}: J^1C\to Q=\Pi_-\subset \Pi,\nonumber\\
 &&p^{(\mu\lambda)}_m\circ\wh L_{YM}=0, \label{5.2a}\\
&&p^{[\mu\lambda]}_m\circ\wh
L_{YM}=\ve^{-2}a^G_{mn}g^{\lambda\alpha}g^{\mu\beta}
\cF^n_{\alpha\beta}\sqrt{\mid g\mid}. \label{5.2b}
\eea

A glance at these expressions shows
that $C_+ = \Ker\wh L$ and $\Pi_- = Q$. It
follows that the splittings (\ref{N31}) and (\ref{N32}) are similar the
splittings (\ref{N18}) and (\ref{N20}).
Therefore, to construct the complete family of multimomentum Hamiltonian forms
associated with the Yang-Mills Lagrangian density (\ref{5.1}), we can follow
the general procedure from previous Section.

Let us consider
connections  (\ref{N16}) on the  bundle $C$  of principal connections which
take their values into $\Ker\wh L$:
 \begin{equation}
  S:C\to C_+, \qquad
S^m_{\m\la}-S^m_{\la\m}-c^m_{nl}k^n_\la k^l_\m=0. \label{69}
\end{equation}
For all these connections, the
 Hamiltonian forms
\ben
 &&H=p^{\mu\lambda}_mdk^m_\mu\wedge\omega_\lambda-
p^{\mu\lambda}_mS^m_{\mu\lambda}\omega-\wt{\cH}_{YM}\omega, \label{5.3}\\
&&\wt{\cH}_{YM}= \frac{\ve^2}{4}a^{mn}_Gg_{\mu\nu}
g_{\lambda\beta} p^{[\mu\lambda]}_m p^{[\nu\beta]}_n\mid g\mid ^{-1/2},
\nonumber
 \een
are associated with the Lagrangian density $L_{YM}$ and constitute the
complete family.
Moreover, we can minimize this complete family if we restrict our
consideration to connections (\ref{69}) of the following type.
Given a symmetric linear connection $K$
 on the cotangent bundle $T^*X$ of $X$,  every principal connection $B$ on
$P$ is lifted to  the  connection $S_B$ (\ref{69})
such that
 \[
S_B\circ B={\cal S}\circ J^1B,
\]
\begin{equation}
 S_B{}^m_{\mu\lambda}=\frac{1}{2} [c^m_{nl}k^n_\lambda
k^l_\mu  +\dr_\mu B^m_\lambda+\dr_\lambda B^m_\mu -c^m_{nl}
(k^n_\mu B^l_\lambda+k^n_\lambda B^l_\mu)] -
K^\beta{}_{\mu\lambda}(B^m_\beta-k^m_\beta). \label{3.7}
\end {equation}
We denote the  Hamiltonian form	 (\ref{5.3}) for the
connections $S_B$ (\ref{3.7})	 by $H_B$.

 The corresponding Hamilton equations
for  sections $r$ of $\Pi\to X$ read
 \ben
 &&\dr_\lambda p^{\mu\lambda}_m=-c^n_{lm}k^l_\nu
p^{[\mu\nu]}_n+c^n_{ml}B^l_\nu p^{(\mu\nu)}_n
-K^\mu{}_{\lambda\nu}p^{(\lambda\nu)}_m,
\label{5.5} \\
&&\dr_\lambda k^m_\mu+ \dr_\mu
k^m_\lambda=2S_B{}^m_{(\mu\lambda)}\label{5.6}
\een
plus the equation (\ref{5.2b}).

The equations (\ref{5.6}) and (\ref{5.2b})
are similar to the equations (\ref{N23}) and (\ref{N28}) respectively. The
equations (\ref{5.2b}) and (\ref{5.5}) restricted to the constraint space
(\ref{5.2a}) are the familiar Yang-Mills equations for $A=\p_{\Pi C}\circ r.$
These equations are the same
for all  Hamiltonian forms $H_B$ and exemplify the
Hamilton-De Donder  equations (\ref{N46}). Different
Hamiltonian forms $H_B$ lead to different   equations (\ref{5.6}) which
play the role of the gauge-type condition
\[
S_B{}\circ A={\cal S}\circ J^1A.
\]
At the same time, given a solution $A$ of the Yang-Mills equations,
 there always exists a multimomentum Hamiltonian form $H_B$ such that
 \[
\wh H_B\circ\wh L_{YM}\circ J^1A=J^1A.
\]
 For instance, this is
$H_{B=A}.$ It follows that the  Hamiltonian forms $H_B$
constitute a complete family.

The gauge-type conditions (\ref{5.6}) however differ from the familiar
gauge conditions in gauge theory.
 Gauge-type classes introduced in previous Section are
preimages ${\cF}_{-1}(e)$ of points $e\in C_-$, whereas the orbits of the
gauge group acting on $J^1C$ belong to preimages $({\rm pr}_1\circ\cF)_{-1}
(v)$ of points
\[
v\in \op\w^2 T^*X\op\otimes_X V^GP.
\]
The gauge-type
 conditions (\ref{5.6}) are universal conditions on sections of the
jet bundle $J^1C\to C$, whereas the familiar gauge conditions are locally
universal conditions on sections of $J^1C\to X$ and single out a
representative in each class of gauge conjugate potentials
 (with accuracy to the Gribov ambiguity).
Namely, if  $A$ is a solution of the Yang-Mills
equations, there exists a gauge conjugate $A'$ which also is a
solution of the same Yang-Mills equations and satisfies given gauge
conditions. However, not every solution of the Yang-Mills equations
obeys one or another
gauge condition utilized in gauge theory. In
this sense, the system of these gauge conditions is not complete. In gauge
theories, this deficiency is not important since all conjugate gauge
potentials are treated as physically equivalent, otherwise for other
degenerate field systems, e.g. the Proca field.
At the same time, we have a complete family of gauge-type conditions.
 For instance, in cases of electromagnetic fields and the Proca field, the
gauge-type condition  (\ref{5.6}) takes the form
\[
\dr_\m A_\nu + \dr_\nu A_\m = \al_{\m\nu}(x).
\]
In particular, one can
 reproduce the standard Lorentz and $\al$-gauge conditions or introduce
other conditions on
$\dr_\m A_\nu + \dr_\nu A_\m$.

\section{Gravity in multimometum variables}

 In this Section, $X$ is a 4-dimensional world manifold  which obeys the
well-known topological conditions in order that a gravitational field
exists on $X^4$.

In the gauge gravitation theory, classical gravity is described by pairs of
 tetrad fields and  reducible Lorentz connections  which
play the role of  gauge gravitational potentials \cite{3sar}.
In the absence of Dirac fermion fields,
one can follow the affine-metric formulation of General Relativity when
gravitational variables are both a pseudo-Riemannian metric $g$ on a world
manifold $X^4$ and a linear connection $K$  on $TX$. We
call them a world metric and a world connection.

The world connections are associated with  principal connections on the
principal bundle $LX\to X^4$ of linear frames in  $TX$.
The structure group of $LX$ is $ GL_4=GL^+(4,{\bf R}).$
 Hence, there is the 1:1 correspondence between the
world connections  and the global sections of the bundle of principal
connections
\[
C=J^1LX/GL_4.
\]

 There is the 1:1 correspondence between the world metrics $g$ on
$X^4$  and the global sections of the  bundle $\Sigma$ of
pseudo-Euclidean bilinear forms in tangent spaces to $X^4$. This bundle is
associated with $LX$. The 2-fold covering of the bundle $\Si$
is the quotient bundle $LX/SO(3,1)$.

Thus, the configuration space of the
affine-metric gravitational variables is represented by the
 product of the corresponding jet manifolds
\begin{equation}
J^1C\op\times_{X^4}J^1\Sigma. \label{W2}
\end{equation}
It is provided   with the adapted coordinates
$(x^\mu, g^{\alpha\beta}, k^\alpha{}_{\beta\mu}, g^{\alpha\beta}{}_\lambda,
k^\alpha{}_{\beta\mu\lambda})$ with respect
to a holonomic bundle atlas of $LX$.

Also the phase space $\Pi$ is the product of the Legendre manifolds
(\ref{00}) of the bundles $C$ and $\Si$.
 The fibred coordinates of $\Pi$  are
$(x^\mu, g^{\alpha\beta}, k^\alpha{}_{\beta\mu},
p_{\alpha\beta}{}^\lambda, p_\alpha{}^{\beta\mu\lambda}).$

On the configuration space (\ref{W2}), the  Hilbert-Einstein  Lagrangian
density of General Relativity reads
\begin{equation}
L_{HE}=-\frac{1}{2\kappa}g^{\beta\lambda}\cF^\alpha{}_{\beta\alpha\lambda}
\sqrt{-g}\omega,\label{5.17}
\end{equation}
 \[
\cF^\alpha{}_{\beta\nu\lambda}=k^\alpha{}_{\beta\lambda\nu}-
k^\alpha{}_{\beta\nu\lambda}+k^\alpha{}_{\varepsilon\nu}
k^\varepsilon{}_{\beta\lambda}-k^\alpha{}_{\varepsilon\lambda}
k^\varepsilon{}_{\beta\nu}.
\]
It is affine, and we can follow the general procedure
>from Section 6. The corresponding Legendre morphism is given by the
expressions
  \ben
&& p_{\alpha\beta}{}^\lambda\circ \wh L_{HE}=0,\nonumber
\\ &&   p_\alpha{}^{\beta\nu\lambda}\circ \wh L_{HE}
=\pi_\alpha{}^{\beta\nu\lambda} =\frac{1}{2\kappa}(\delta^\nu_\alpha
g^{\beta\lambda}-\delta^\lambda_\alpha g^{\beta\nu})\sqrt{-g}.\label{5.18}
\een

Let us construct complete family of  Hamiltonian
forms (\ref{N26}) associated with the Lagrangian density (\ref{5.17}). To
minimize it, we consider a certain subset of connections on the bundle
$C\times\Si$. Let $K$ be a world connection and
\be
&& S_K{}^\alpha{}_{\beta\nu\lambda}=\frac12
[k^\alpha{}_{\varepsilon\nu}
k^\varepsilon{}_{\beta\lambda}-k^\alpha{}_{\varepsilon\lambda}
k^\varepsilon{}_{\beta\nu} +\dr_\lambda K^\alpha{}_{\beta\nu}
+\dr_\nu K^\alpha{}_{\beta\lambda}\\
&& -2K^\varepsilon{}_{(\nu\lambda)}(K^\alpha{}_{\beta\varepsilon}
-k^\alpha{}_{\beta\varepsilon}) +
K^\varepsilon{}_{\beta\lambda}k^\alpha{}_{\varepsilon\nu}
+K^\varepsilon{}_{\beta\nu}k^\alpha{}_{\varepsilon\lambda} -
K^\alpha{}_{\varepsilon\lambda}k^\varepsilon{}_{\beta\nu}
-K^\alpha{}_{\varepsilon\nu}k^\varepsilon{}_{\beta\lambda}]
\ee
the corresponding connection (\ref{3.7}) on the bundle $C$. Let $K'$ be
another symmetric world connection which induces a connection on the bundle
$\Si$.  On the bundle $C\times\Si$, we consider the following connection
\begin{equation}
 \G^{\al\bt}{}_\la =-{K'}^\alpha{}_{\varepsilon\lambda}
g^{\varepsilon\beta} -
{K'}^\beta{}_{\varepsilon\lambda} g^{\alpha\varepsilon}, \qquad
\Gamma^\alpha{}_{\beta\nu\lambda} = S_K{}^\alpha{}_{\beta\nu\lambda}
-R^\alpha{}_{\beta\nu\lambda} \label{N34}
 \end{equation}
  where	 $R$ is the curvature of the connection $K$. The corresponding
 Hamiltonian form (\ref{N26}) is given by the expression
 \ben
&& H_{HE}=(p_{\alpha\beta}{}^\lambda dg^{\alpha\beta} +
p_\alpha{}^{\beta\nu\lambda}dk^\alpha{}_{\beta\nu})\wedge\omega_\lambda
-\cH_{HE}\omega, \nonumber \\
&& \cH_{HE}=-p_{\alpha\beta}{}^\lambda({K'}^\alpha{}_{\varepsilon\lambda}
g^{\varepsilon\beta} +{K'}^\beta{}_{\varepsilon\lambda}
g^{\alpha\varepsilon})
+p_\alpha{}^{\beta\nu\lambda}\Gamma^\alpha{}_{\beta\nu\lambda}
 -R^\alpha{}_{\beta\nu\lambda}
(p_\alpha{}^{\beta\nu\lambda}-\pi_\alpha{}^{\beta\nu\lambda}). \label{5.19}
\een
It is associated with the Lagrangian density $L_{HE}$.
Given $H_{HE}$ (\ref{5.19}),
the corresponding covariant Hamilton equations for General Relativity read
 \bea
 &&\dr_\lambda
g^{\alpha\beta} +{K'}^\alpha{}_{\varepsilon\lambda}g^{\varepsilon\beta}
+{K'}^\beta{}_{\varepsilon\lambda}g^{\alpha\varepsilon}=0, \label{5.20a} \\
&&\dr_\lambda k^\alpha{}_{\beta\nu}=\Gamma^\alpha{}_{\beta\nu\lambda}
-R^\alpha{}_{\beta\nu\lambda}, \label{5.20b} \\
&&\dr_\lambda p_{\alpha\beta}{}^\lambda =p_{\varepsilon\beta}{}^\sigma
{K'}^\varepsilon{}_{\alpha\sigma} +
p_{\varepsilon\alpha}{}^\sigma {K'}^\varepsilon{}_{\beta\sigma}
+\frac{1}{\kappa}(R_{\alpha\beta} -\frac12g_{\alpha\beta}R)\sqrt{-g},
\label{5.20c} \\
 && \dr_\lambda p_\alpha{}^{\beta\nu\lambda}=
 -p_\alpha{}^{\varepsilon[\nu\gamma]}k^\beta{}_{\varepsilon\gamma}
 +p_\varepsilon{}^{\beta[\nu\gamma]}
k^\varepsilon{}_{\alpha\gamma} \nonumber \\
&& \qquad - p_\alpha{}^{\beta\varepsilon\gamma}
K^\nu{}_{(\varepsilon\gamma)} -p_\alpha{}^{\varepsilon(\nu\gamma)}
K^\beta{}_{\varepsilon\gamma}  +p_\varepsilon{}^{\beta(\nu\gamma)}
K^\varepsilon{}_{\alpha\gamma}. \label{5.20d}
\eea

The equations (\ref{5.20a}) and (\ref{5.20b}) represent the gauge-type
 conditions
(\ref{N27}). In accordance with the canonical splitting (\ref{N31}), the
equations (\ref{5.20b}) are brought into the pair of equations
\ben
&&\cF^\alpha{}_{\beta\nu\lambda}=R^\alpha{}_{\beta\nu\lambda}, \label{5.21}\\
 && \dr_\nu(K^\alpha{}_{\beta\lambda} -k^\alpha{}_{\beta\lambda})
+\dr_\lambda(K^\alpha{}_{\beta\nu} -k^\alpha{}_{\beta\nu})
  -2K^\varepsilon{}_{(\nu\lambda)}
(K^\alpha{}_{\beta\varepsilon} -k^\alpha{}_{\beta\varepsilon}) +
K^\varepsilon{}_{\beta\lambda}k^\alpha{}_{\varepsilon\nu}\nonumber\\
&& \qquad +K^\varepsilon{}_{\beta\nu}k^\alpha{}_{\varepsilon\lambda}
  -K^\alpha{}_{\varepsilon\lambda}k^\varepsilon{}_{\beta\nu}
-K^\alpha{}_{\varepsilon\nu}k^\varepsilon{}_{\beta\lambda}=0.
\label{5.22}
 \een
For a given world metric $g$ and a world connection $k$, there
always exist connections $K'$ and $K$ such that these gauge conditions hold.
It follows that the  Hamiltonian forms (\ref{5.19}) consitute a
complete family.

 Being restricted to the constraint space (\ref{5.18}), the equations
(\ref{5.20c}) and (\ref{5.20d}) read
\ben
&& \frac{1}{\kappa}(R_{\alpha\beta}
-\frac12 g_{\alpha\beta}R)\sqrt{-g} =0, \label{5.23} \\
&& D_\al(\sqrt{-g}g^{\nu\bt}) - \delta^\nu_\al
D_\la(\sqrt{-g}g^{\la\bt}) +\sqrt{-g}[g^{\nu\bt}(k^\la{}_{\al\la} -
k^\la{}_{\la\al}) \nonumber\\
&& \qquad + g^{\la\bt}(k^\nu{}_{\la\al} - k^\nu{}_{\al\la}) + \delta^\nu_\al
g^{\la\bt} (k^\m{}_{\m\la} - k^\m{}_{\la\m})] =0, \label{5.24}
 \een
\[
D_\la g^{\al\bt}= \dr_\la g^{\al\bt} + k^\al{}_{\m\la}g^{\m\bt} +
k^\bt{}_{\m\la}g^{\al\m}.
\]
 Substituting the condition
(\ref{5.21}) into the equation (\ref{5.23}), we obtain the Einstein equations
\begin{equation}
 \cF_{\alpha\beta}-\frac12 g_{\alpha\beta}\cF= 0.\label{5.25}
\end{equation}
The equations (\ref{5.24}) and (\ref{5.25}) are the familiar equations for a
gravity in the nonsymmetric Palatini variables. The former is the equation for
torsion and nonmetricity terms of the connection $k^\al{}_{\bt\nu}$. In the
absence of matter, it has the  well-known solution
\[
k^\al{}_{\bt\nu} =\{^\al{}_{\bt\nu}\} - \frac12\delta^\al_\nu V_\bt,
\qquad D_\al g^{\bt\g}= V_\al g^{\bt\g},
\]
where $V_\al$ is an arbitrary covector field corresponding to the
so-called projective freedom.
\medskip

The multimomentum quantum field theory has been hampered by the lack of
satisfactory commutation relations between multimomentum canonical
variables. At the same time, the multimomentum Hamiltonian formalism
can be extended to quantum field theory if one considers chronological
forms, but not commutation relations \cite{5sard}. Moreover, it
may incorporate together the canonical and algebraic approaches to
quantization of fields.

\end{document}